\documentclass[letterpaper, 10 pt, conference]{ieeeconf} 

\usepackage{amssymb,latexsym,color,amsmath,pifont,epsfig,graphicx}

\usepackage{amsthm}
\usepackage{setspace,cite}
\usepackage{subfig}  
\usepackage{algorithm}
\usepackage{algorithmic}
\usepackage{amssymb}
\usepackage{amsmath}
\usepackage[font=scriptsize]{caption}
\usepackage{cite}
\usepackage{soul}
\usepackage{gensymb}
\usepackage{graphics}
\usepackage{multirow}
\usepackage{bm}
\usepackage{rotating}
\newtheorem{theorem}{Theorem}
\usepackage{epsfig}
\usepackage{amssymb,latexsym,color,amsmath,pifont,epsfig,mathtools}
\usepackage{graphicx,ctable,booktabs}
\usepackage{indentfirst}
\usepackage{dsfont}
\usepackage{bbm}
\usepackage{amsgen}
\usepackage[outdir=./]{epstopdf}
\usepackage{url}

\theoremstyle{plain}
\providecommand{\lemmaname}{Lemma}

\allowdisplaybreaks
\IEEEoverridecommandlockouts

\begin{document}

\title{\bf Reduced-order Aggregate Dynamical Model for Wind Farms}

\author{Sanjana Vijayshankar, Victor Purba, Peter J. Seiler, and Sairaj V. Dhople%
\thanks{Sanjana Vijayshankar, Victor Purba, and Sairaj V. Dhople are with the Department of Electrical and Computer Engineering, University of Minnesota, Minneapolis, MN 55455. E-mails: \{vijay092, purba002, sdhople\}@umn.edu. Peter J. Seiler is with the Department of Aerospace and Engineering Mechanics, University of Minnesota, Minneapolis, MN 55455. E-mail: seile017@umn.edu.}%
}

\maketitle

\begin{abstract}
This paper presents an aggregate reduced-order model  for a wind farm composed of identical parallel-connected Type-3 wind turbines. The model for individual turbines includes mechanical dynamics (arising from the turbine and doubly fed induction generator) and electrical dynamics (arising from the rotor-side and grid-side converters and associated filters). The proposed aggregate wind-farm model is structure preserving, in the sense that the parameters of the model are derived by scaling corresponding ones from the individual turbines. The aggregate model hence maps to an equivalent--albeit fictitious--wind turbine that captures the dynamics corresponding to the entire wind farm. The reduced-order model has obvious computational advantages, but more importantly, the presented analysis rigorously formalizes parametric scalings for aggregate wind-turbine models that have been applied with limited justification in prior works. Exhaustive numerical simulations validate the accuracy and computational benefits of the proposed reduced-order model.

\end{abstract}
\section{Introduction}
The largest wind farm in the US is located in Altamont Pass, CA, and it has $4800$ wind turbines. Modeling and analysis of large wind farms---particularly, focusing on their aggregate behavior and dynamic interactions with the remainder of the bulk power system---hinges on the availability of computationally scalable and accurate dynamical models. Taking a step in this direction, this paper proposes a reduced-order aggregate model for a wind farm built with an arbitrary number of parallel-connected Type-3 wind turbines~\cite{burton2011wind,manwell2010wind,hiskens2012dynamics}. The aggregate model is \emph{structure preserving} in the sense that it maps to an equivalent (larger and fictitious) wind turbine with systematically scaled parameters and ratings. Furthermore, while the setup with identical turbines and a parallel collector system is admittedly idealistic, the reduced-order model pre
serves all nonlinearities in the individual turbine model, and therefore \emph{exactly} captures the input-output behavior of the wind farm. 

\begin{figure}[b!]
	\centering
     \includegraphics[width=0.4\textwidth]{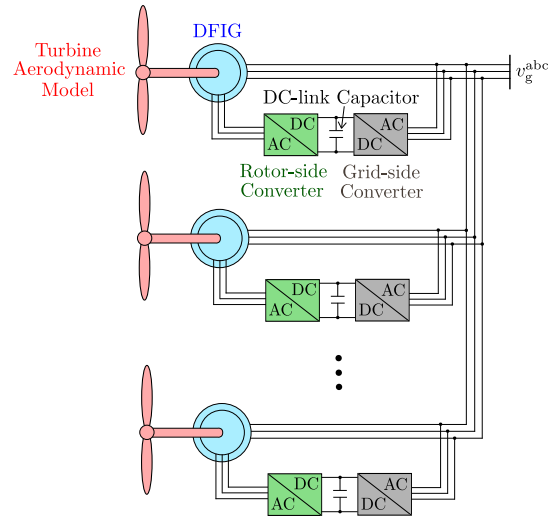}
   \caption{Parallel connection of $N$ identical Type-3 wind turbines. We show that the electrical and mechanical dynamics of this system can be captured by the model sketched in Fig.~\ref{fig:blockdiagram} with scaled parameters depending on the number of wind turbines (highlighted in red).}
   \label{fig:parallel}
\end{figure}

\begin{figure*}[!t]
	\centering
     \includegraphics[width=0.8\textwidth]{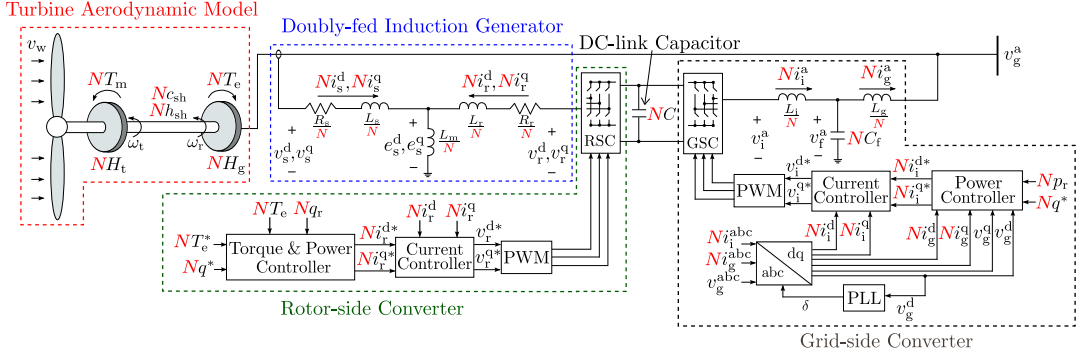}
   \caption{Block diagram depicting controllers, equivalent circuit representations, and turbine aerodynamics for Type-3 wind turbines (one electrical phase is depicted for simplicity). The aggregate model for a collection of $N$ parallel-connected turbines---shown in Fig.~\ref{fig:parallel}---is obtained with highlighted parametric scalings. (Resulting scalings in pertinent state variables are also highlighted accordingly.) Blocks labeled RSC and GSC correspond to the switching power-semiconductor devices in the rotor-side and grid-side converters, respectively; and the blocks labeled PWM generate the pulse width modulation signals to switch these devices.}
   \label{fig:blockdiagram}
\end{figure*}

Subsystems of Type-3 wind turbines include: i)~a turbine aerodynamic model that captures the conversion of kinetic energy in the wind to low-speed rotational energy, ii)~a doubly-fed induction generator (DFIG) that converts mechanical energy to electrical energy, iii)~a rotor side converter for maximum power-point tracking, and iv)~a grid-side converter that maintains a constant DC-link voltage. (See Fig. \ref{fig:parallel} for an illustration.) Adopting commonly used models for these subsystems from the literature, the dynamical model for an individual turbine examined here has $27$ states. Our proposed aggregate reduced-order model for the
wind farm systematically establishes parametric scalings (in electrical and mechanical subsystems) such that the input-output behavior is perfectly captured by a dynamical model that has the same structure and dimension as any individual turbine. In particular, the reduced-order model for the wind farm also has $27$ states. The idea is illustrated by the block diagram sketched in Fig.~\ref{fig:blockdiagram}. (Setting $N=1$ recovers the model for an individual turbine, while an arbitrary value of $N$ captures the dynamics for the system in Fig.~\ref{fig:parallel}. Blocks shaded or enclosed in dashed lines with the same color across the two figures represent the same subsystem.)

Parametric scalings relating individual and aggregate models have been presented in the literature, see e.g.,~\cite{zou2014survey,al2016aggregated, duraisamy2017dynamic}. However, these works do not provide rigorous analysis to validate that the aggregate model synthesized with the scaled parameters faithfully captures the dynamics of the collection of individual turbines pointwise in time. Literature pertinent to the present work also includes a variety of methods that have been applied for model reduction for wind energy conversion systems~\cite{slootweg2003aggregated, conroy2009aggregate, kim2015dynamic,brochu2011validation, gautam2009impact,pulgar2011towards, annoni2015low, ghosh2014reduced,al2016aggregated}. In such approaches, a linear time-invariant system is first obtained from the originating nonlinear dynamical system and then model-order reduction methods developed for linear systems are employed. Compared to these methods, in this work we retain all the nonlinearities from the original model in the aggregate reduced-order model. The main motivation for retaining the nonlinearities is that turbine dynamics are a function of wind speed, and therefore, linearized representations obtained at fixed wind speeds  fail to capture the dynamics of the aggregation when wind conditions are varying. 

A key application of reduced-order models is in power-system dynamic studies. Given the lack of high-fidelity farm-level dynamic models, in many instances, e.g.,~\cite{pereira2014high,muljadi2011short}, the turbines are modeled simply as current injections and all turbine dynamics are neglected. With such simplified models, potential instabilities caused in the power electronic circuits and the power system network cannot be captured. The aggregate model we present preserves all electromechanical dynamics and can be used to investigate phenomena of interest in the electrical and mechanical domains with limited computational burden. As another potential application, in ~\cite{shapiro2017model,katic1987simple,8430751,6626635}, wake models are derived to capture the effect of aerodynamic interactions between individual turbines. Such models consider a regular rectangular wind farm with each row aligned perpendicular to the prevailing wind direction and turbines in every row assumed to receive the same wind speed. However, in this setup, the dynamics of turbines, both mechanical and electrical, are typically ignored. We anticipate that an augmented model developed with: i)~the aggregate wind farm model proposed in this paper, and the ii)~wake models proposed in, e.g.,~\cite{shapiro2017model,katic1987simple,8430751,6626635} will better describe the time-varying impact of changing turbine kinetic energy extraction on the total wind farm power production. Finally, we note that time-scale separation arguments can be applied to reduce the order of the aggregate wind-farm dynamical model to exclusively examine electrical (faster) or mechanical (slower) phenomena.

Admittedly, the architecture we analyze abstracts away details including: flow dynamics within the farms leading to variability in wind speeds for different turbines, differences in hardware and control parameters, and interconnecting impedances in the wind farm distribution network. While these are a focus of ongoing work, we believe the present paper provides a meaningful contribution to the literature since it rigorously establishes parametric scalings needed to obtain equivalent wind-turbine representations for large wind farms, in the process formalizing modeling assumptions that have been taken for granted previously. Furthermore, as discussed above, it establishes an electromechanical dynamical model critical for accurate representation of wind-farm dynamics in bulk power system studies and wind-farm wake studies.

The remainder of this paper is organized as follows. Section~\ref{sec:Model} describes the dynamical equations of the wind turbine system. The main result pertaining to establishing the reduced-order aggregate model is presented in Section~\ref{sec: ROM}. Numerical simulations to validate the reduced-order model are provided in Section~\ref{sec: Results}. Finally, we conclude with a few directions for further work in Section~\ref{sec:Conclusions}.

\section{Model Description}
\label{sec:Model}
In this section, we outline the state-space model for a single turbine by presenting the dynamics for the following: turbine aerodynamics, doubly-fed induction generator (DFIG), rotor-side converter (with associated controllers), grid-side converter (with output $LCL$-filter and associated controllers), and DC-link capacitor. We have made every effort to define model parameters in a self-contained fashion in Appendix~B and a subset are illustrated in Fig.~\ref{fig:blockdiagram}. Parameters that are not explicitly defined or illustrated are assumed to be contextually obvious. 

With regard to electrical variables, we assume balanced three-phase operation. Balanced three-phase signals ($x^\mathrm{a}, x^\mathrm{b}, x^\mathrm{c}$) are transformed into equivalent DC signals ($x^\mathrm{q}, x^\mathrm{d}$) using Park's transformation:
\begin{align*}
\begin{bmatrix} x^\mathrm{q} \\ x^\mathrm{d} \end{bmatrix}
&=
\frac{2}{3}
\begin{bmatrix}
\cos(\delta) & \cos(\delta - \frac{2 \pi}{3}) & \cos(\delta + \frac{2 \pi}{3}) \\[0.1cm]
-\sin(\delta) & -\sin(\delta - \frac{2 \pi}{3}) & -\sin(\delta + \frac{2 \pi}{3})
\end{bmatrix}
\begin{bmatrix} x^\mathrm{a} \\ x^\mathrm{b} \\ x^\mathrm{c} \end{bmatrix},
\end{align*}
where $\delta$ is the angle generated by the PLL. 


\subsection{Turbine Aerodynamic Model}
The turbine aerodynamic model derives from a two-mass model and captures the dynamics of the generator speed, twist angle, and turbine speed: 
\begin{align}
\dot{\omega}_\mathrm{r} &= \frac{1}{2H_\mathrm{g}}(k_\mathrm{sh} \theta_\mathrm {tw} + c_\mathrm{sh}\omega_\mathrm{nom} (\omega_\mathrm t -\omega_\mathrm r) -T_\mathrm e),\label{eq:aerodynamics1} \\
\dot{\theta}_\mathrm{tw} &= \omega_\mathrm{nom} (\omega_\mathrm{t} - \omega_\mathrm{r}), \label{eq:aerodynamics2} \\
\dot{\omega}_\mathrm{t} &= \frac{1}{2H_\mathrm{t}}(T_\mathrm{m}-k_\mathrm{sh} \theta_\mathrm{tw} - c_\mathrm{sh}\omega_\mathrm{nom} (\omega_\mathrm{t} -\omega_\mathrm{r})) \label{eq:aerodynamics3}.
\end{align}
Above, $\omega_\mathrm{t}, \omega_\mathrm{r}$ denote turbine and generator speed,  $\theta_\mathrm{tw}$ denotes the equivalent twist angle of the drive shaft, and  $T_\mathrm e$ denotes the electromagnetic torque of the generator (it will be formally defined in Section~\ref{subsec:generator}). Finally, $T_\mathrm{m}$ denotes mechanical torque of the turbine, and it is given by
\begin{equation}
T_\mathrm{m} = \frac{1}{2T_{\mathrm{m,base}}}\left(\rho \pi R^2 C_p(\lambda,\beta)\frac{v_{\mathrm{w}}^3}{\omega_\mathrm{t}}\right),
\end{equation}
where $C_p(\lambda,\beta)$ is the turbine performance coefficient, $v_{\mathrm{w}}$ is the wind speed, $\lambda = \frac{\omega_\mathrm{t} R}{v_{\mathrm{w}}}$  is the tip-speed ratio, $\beta$ is the blade-pitch angle and $T_{\mathrm{m,base}}$ is the turbine base torque.

\subsection{Doubly-fed Induction Generator (DFIG)}
\label{subsec:generator}
The states of the DFIG are the stator currents $(i_\mathrm{s}^\mathrm{q}, i_\mathrm{s}^\mathrm{d})$ and rotor transient voltages $(e_\mathrm{s}^\mathrm{q}, e_\mathrm{s}^\mathrm{d})$. The dynamics of the generator in the $\mathrm{dq}$-reference frame correspond to those of the equivalent circuit enclosed in dashed blue lines (for $N=1$) in Fig.~\ref{fig:blockdiagram}, and given by~\cite{mei2008modelling}
\begin{align}
\dot{i}_\mathrm{s}^\mathrm{q} &= \frac{\omega_\mathrm{nom}}{L_{\mathrm s}'}\big( -R_1i_\mathrm{s}^\mathrm{q} + \omega_\mathrm{s} L_{\mathrm s}'i_\mathrm{s}^\mathrm{d} + \frac{\omega_\mathrm r}{\omega_\mathrm s} \mathrm{e}_\mathrm{s} ^ \mathrm{q} -\frac{1}{T_\mathrm{r}\omega_\mathrm s} \mathrm{e}_\mathrm{s} ^ \mathrm{d}  \nonumber \\
&\quad  - v_\mathrm{g}^{\mathrm{q}} + K_\mathrm{mrr}v_\mathrm{r}^\mathrm{q} \big), \nonumber  \\
\dot{i}_\mathrm{s}^\mathrm{d} &= \frac{\omega_\mathrm{nom}}{L_{\mathrm s}'} \big( -R_1i_\mathrm{s}^\mathrm{d} - \omega_\mathrm{s} L_{\mathrm s}'i_\mathrm{s}^\mathrm{q} + \frac{\omega_\mathrm r}{\omega_\mathrm s} \mathrm{e}_\mathrm{s} ^ \mathrm{d} +\frac{1}{T_\mathrm{r}\omega_\mathrm s} \mathrm{e}_\mathrm{s} ^ \mathrm{q} \\
&\quad - v_\mathrm{g}^{\mathrm{d}} + K_\mathrm{mrr}v_\mathrm{r}^\mathrm{d} \big) , \nonumber \\
\dot{e}_\mathrm{s}^\mathrm{q} &= \omega_{\mathrm{nom}}\omega_\mathrm{s}( R_2i_\mathrm{s}^\mathrm{d} - \frac{\mathrm{e}_\mathrm{s} ^ \mathrm{q}}{T_\mathrm{r} \omega_\mathrm{s}} + (1- \frac{\omega_\mathrm r}{\omega_\mathrm s})\mathrm{e}_\mathrm{s} ^ \mathrm{d} - K_\mathrm{mrr}v_\mathrm{r}^\mathrm{d}), \nonumber \\
\dot{e}_\mathrm{s}^\mathrm{d} &=- \omega_{\mathrm{nom}}\omega_\mathrm s(R_2i_\mathrm{s}^\mathrm{q} + \frac{\mathrm{e}_\mathrm{s} ^ \mathrm{d}}{T_\mathrm{r} \omega_\mathrm s} + (1- \frac{\omega_\mathrm r}{\omega_\mathrm s})\mathrm{e}_\mathrm{s} ^ \mathrm{q} - K_\mathrm{mrr}v_\mathrm{r}^\mathrm{q}), \nonumber
\end{align}
where $v_\mathrm{r}^\mathrm{d}, v_\mathrm{r}^\mathrm{q}$ denote the rotor voltages in the $\mathrm{dq}$-frame. All quantities mentioned above except $v_\mathrm{r}^\mathrm{d}$ and $ v_\mathrm{r}^\mathrm{q}$  are referred to the stator side. Parameters $L_{\mathrm s}'$, $ T_\mathrm{r}$, $R_1$, and $R_2$  map to the circuit shown in Fig. \ref{fig:blockdiagram} with appropriate scalings reflecting their representation on the stator side. Refer Appendix~B for more details. The outputs of the DFIG are the electromagnetic torque of the generator, $T_\mathrm{e}$, and the total active and reactive powers ($p_\mathrm{tot}$ and $q_\mathrm{tot}$) delivered to the grid:
\begin{align}
T_\mathrm e &= \frac{v_\mathrm{g} ^ \mathrm{q}}{\omega_\mathrm{s}}i_\mathrm{s}^\mathrm{q} + \frac{v_\mathrm{g}^\mathrm{d}}{\omega_\mathrm s}i_\mathrm{s}^\mathrm{d},\\
p_\mathrm{tot} &= p_\mathrm{s} + p_\mathrm{g} = v_\mathrm{g}^\mathrm{d} i_\mathrm{s}^\mathrm{d} + v_\mathrm{g}^\mathrm{q} i_\mathrm{s}^\mathrm{q} + p_\mathrm{g}, \label{eq:p_tot}\\
q_\mathrm{tot} &= q_\mathrm{s} + q_\mathrm{g} = - v_\mathrm{g}^\mathrm{q}i_\mathrm{s}^\mathrm{d} + v_\mathrm{g}^\mathrm{d} i_\mathrm{s}^\mathrm{q} + q_\mathrm{g}, \label{eq:q_tot}
\end{align}
where $p_\mathrm{s}$ and $q_\mathrm{s}$ are the active and reactive power from the stator and $p_\mathrm{g}$ and $q_\mathrm{g}$  are the powers delivered from the grid side converters (see Section~\ref{sec:grid-converter} for details).

\subsection{Rotor-side Converter}
\label{subsec:rotor-converter}
The control architecture of the rotor side converter~\cite{mei2008modelling} consists of outer-loop reactive-power and electromagnetic-torque controllers, and inner-loop current controllers. This is sketched in Fig.~\ref{fig:blockdiagram} in dashed green lines. The rotor-side converter is also interchangeably referred to as the rectifier. 

\subsubsection{Rotor-side reactive-power controller} This is composed of a PI controller with PI gains $k^p_\mathrm{RPC}, k^i_\mathrm{RPC}$. It tracks the reactive power set-point, and as output, yields the reference to the $\mathrm{d}$-component of the current controller:
\begin{equation}
i_\mathrm{r}^\mathrm{d*} = k^p_\mathrm{RPC} (q^* - q_\mathrm{s}) + k^i_\mathrm{RPC} \int (q^* - q_\mathrm{s}),
\end{equation}
where $q_\mathrm{s}$ is the stator reactive power~\eqref{eq:q_tot}.

\subsubsection{Electromagnetic-torque controller} It is composed of a PI controller with PI gains $k^p_\mathrm{RTC}, k^i_\mathrm{RTC}$. It tracks an optimal electromagnetic-torque reference, given by $T_\mathrm{e}^* = K_\mathrm{opt} \omega_\mathrm{r}^2$, where $K_\mathrm{opt}$ denotes the optimum torque constant at a given wind speed. The reference input is thus state-dependent. The output of the torque controller is the reference for the $\mathrm{q}$-component of the current controller:
\begin{equation}
i_\mathrm{r}^\mathrm{q*} = k^p_\mathrm{RTC} (T_\mathrm{e}^* - T_\mathrm{e}) + k^i_\mathrm{RTC} \int (T_\mathrm{e}^* - T_\mathrm{e}).
\end{equation}

\subsubsection{Rotor-side current controller} It consists of two PI controllers with PI gains $k^{p\mathrm{d}}_\mathrm{RCC}$,  $k^{p\mathrm{q}}_\mathrm{RCC}$, $k^{i\mathrm{d}}_\mathrm{RCC}$, $k^{i\mathrm{q}}_\mathrm{RCC}$, and their outputs are the rotor voltage references:
\begin{align}
v_\mathrm{r}^\mathrm{d*} &= k^{p\mathrm{d}}_\mathrm{RCC}(i_\mathrm{r}^\mathrm{d*} - i_\mathrm{r}^\mathrm{d}) + k^{i\mathrm{d}}_\mathrm{RCC} \int (i_\mathrm{r}^\mathrm{d*} - i_\mathrm{r}^\mathrm{d}), \label{eq:vr_d} \\
v_\mathrm{r}^\mathrm{q*} &= k^{p\mathrm{q}}_\mathrm{RCC} (i_\mathrm{r}^\mathrm{q*} - i_\mathrm{r}^\mathrm{q}) + k^{i\mathrm{q}}_\mathrm{RCC} \int (i_\mathrm{r}^\mathrm{q*} - i_\mathrm{r}^\mathrm{q}), \label{eq:vr_q}
\end{align}
where the rotor current is given by
\begin{equation}
i_\mathrm{r}^\mathrm{d} = \frac{e_\mathrm{s}^\mathrm{q}}{X_\mathrm{m}} - K_\mathrm{mrr}i_\mathrm{s}^\mathrm{d},\quad i_\mathrm{r}^\mathrm{q} = \frac{e_\mathrm{s}^\mathrm{d}}{X_\mathrm{m}} - K_\mathrm{mrr}i_\mathrm{s}^\mathrm{q}, \label{eq:ir}
\end{equation}
and $X_\mathrm{m}$ denotes the impedance of the mutual inductance $L_\mathrm{m}$ (see Fig.~\ref{fig:blockdiagram}). Assuming an ideal converter, the terminal voltage at the output of the converter is the same as its reference. Then, the output power of the rotor is given by
\begin{equation}
p_\mathrm{r} = v_\mathrm{r}^\mathrm{d} i_\mathrm{r}^\mathrm{d} + v_\mathrm{r}^\mathrm{q} i_\mathrm{r}^\mathrm{q},\quad q_\mathrm{r} = -v_\mathrm{r}^\mathrm{q} i_\mathrm{r}^\mathrm{d} + v_\mathrm{r}^\mathrm{d} i_\mathrm{r}^\mathrm{q}.
\end{equation}

\subsection{Grid-side Converter}
\label{sec:grid-converter}
This consists of a full bridge inverter with output $LCL$ filter, and the control architecture is composed of a phase-locked loop (PLL), an outer-loop power controller, and an inner-loop current controller. This is sketched in Fig.~\ref{fig:blockdiagram} enclosed in dashed black lines. The grid-side converter is also interchangeably referred to as the inverter.

\subsubsection{Phase-locked loop (PLL)} The PLL synchronizes with the grid by modulating angle $\delta$ so that $v_\mathrm{g}^\mathrm{d}$ diminishes to 0 asymptotically (when $v_\mathrm{g}^\mathrm{d} = 0$, the PLL angle $\delta$ is the same as the instantaneous angle of $v_\mathrm{g}^\mathrm{a}$). It consists of a PI controller with PI gains $k^p_\mathrm{PLL}, k^i_\mathrm{PLL}$ and a low pass filter with cut-off frequency $\omega_{\mathrm{c,PLL}}$. The dynamics of the PLL are:
\begin{align}
\begin{split}
\dot{v}_\mathrm{PLL} &= \omega_\mathrm{c,PLL}(v_\mathrm{g}^\mathrm{d} - v_\mathrm{PLL}),\quad \dot{\phi}_\mathrm{PLL} = -v_\mathrm{PLL}, \\
\dot{\delta} &= 1 - k^p_\mathrm{PLL} v_\mathrm{PLL} + k^i_\mathrm{PLL} \phi_\mathrm{PLL} =: \omega_\mathrm{PLL},
\end{split}
\end{align}
where $v_\mathrm{PLL}$ and $\phi_\mathrm{PLL}$ are internal PLL states.

\subsubsection{Output $LCL$-filter} We consider a lossless $LCL$-filter composed of inverter-side inductor $L_\mathrm{i}$ with current denoted by (in $\mathrm{dq}$-frame) $i_\mathrm{i}^\mathrm{d}, i_\mathrm{i}^\mathrm{q}$, capacitor $C_\mathrm{f}$ with voltage $v_\mathrm{f}^\mathrm{d}, v_\mathrm{f}^\mathrm{q}$, and grid-side inductor $L_\mathrm{g}$ with current $i_\mathrm{g}^\mathrm{d}, i_\mathrm{g}^\mathrm{q}$. The filter is depicted in Fig.~\ref{fig:blockdiagram} at the output of the switching block labeled GSC, and its dynamics are given by
\begin{align}
\dot{i}_\mathrm{i}^\mathrm{dq} &= \frac{\omega_\mathrm{nom}}{L_\mathrm{i}}\left( v_\mathrm{i}^\mathrm{dq} - v_\mathrm{f}^\mathrm{dq}\right) + \begin{bmatrix} 0 & 1 \\ -1 & 0
\end{bmatrix} \omega_\mathrm{nom} \omega_\mathrm{PLL} i_\mathrm{i}^\mathrm{dq}, \\
\dot{i}_\mathrm{g}^\mathrm{dq} &= \frac{\omega_\mathrm{nom}}{L_\mathrm{g}}\left( v_\mathrm{f}^\mathrm{dq} - v_\mathrm{g}^\mathrm{dq}\right) + \begin{bmatrix} 0 & 1 \\ -1 & 0
\end{bmatrix} \omega_\mathrm{nom} \omega_\mathrm{PLL} i_\mathrm{g}^\mathrm{dq}, \\
\dot{v}_\mathrm{f}^\mathrm{dq} &= \frac{\omega_\mathrm{nom}}{C_\mathrm{f}}\left( i_\mathrm{i}^\mathrm{dq} - i_\mathrm{g}^\mathrm{dq}\right) + \begin{bmatrix} 0 & 1 \\ -1 & 0
\end{bmatrix} \omega_\mathrm{nom} \omega_\mathrm{PLL} v_\mathrm{f}^\mathrm{dq},
\end{align}
where $i_\mathrm{i}^\mathrm{dq} := [i_\mathrm{i}^\mathrm{d}, i_\mathrm{i}^\mathrm{q}]^\mathrm{T}$, $i_\mathrm{g}^\mathrm{dq} := [i_\mathrm{g}^\mathrm{d}, i_\mathrm{g}^\mathrm{q}]^\mathrm{T}$, $v_\mathrm{f}^\mathrm{dq} := [v_\mathrm{f}^\mathrm{d}, v_\mathrm{f}^\mathrm{q}]^\mathrm{T}$, $v_\mathrm{i}^\mathrm{dq} := [v_\mathrm{i}^\mathrm{d}, v_\mathrm{i}^\mathrm{q}]^\mathrm{T}$, and $v_\mathrm{g}^\mathrm{dq} := [v_\mathrm{g}^\mathrm{d}, v_\mathrm{g}^\mathrm{q}]^\mathrm{T}$.

\subsubsection{Grid-side power controller} It consists of two low-pass filters with cut-off frequency $\omega_\mathrm{c,PC}$ and two PI controllers with PI gains $k^p_\mathrm{GPC}, k^i_\mathrm{GPC}$. It tracks the reference active and reactive power, given by the rotor active power $p_\mathrm{r}$ and reactive-power setpoint $q^*$, respectively, and it regulates the output power of the grid-side converter. The instantaneous active and reactive power injected to the grid from the grid-side converter are given by
\begin{equation}
p_\mathrm{g} = v_\mathrm{g}^\mathrm{d} i_\mathrm{g}^\mathrm{d} + v_\mathrm{g}^\mathrm{q} i_\mathrm{g}^\mathrm{q},\quad
q_\mathrm{g} = -v_\mathrm{g}^\mathrm{q} i_\mathrm{g}^\mathrm{d} + v_\mathrm{g}^\mathrm{d} i_\mathrm{g}^\mathrm{q}. \label{eq:rotor_power}
\end{equation}
The dynamics of the low-pass filters are:
\begin{equation}
\dot{p}_\mathrm{avg} = \omega_\mathrm{c,PC}(p_\mathrm{g} - p_\mathrm{avg}),\quad
\dot{q}_\mathrm{avg} = \omega_\mathrm{c,PC}(q_\mathrm{g} - q_\mathrm{avg}).
\end{equation}
Outputs are references to the current controller:
\begin{align}
i_\mathrm{i}^\mathrm{d*} &= k^p_\mathrm{GPC} (q^* - q_\mathrm{avg}) + k^i_\mathrm{GPC} \int (q^* - q_\mathrm{avg}), \\
i_\mathrm{i}^\mathrm{q*} &= k^p_\mathrm{GPC} (p_\mathrm{r} - p_\mathrm{avg}) + k^i_\mathrm{GPC} \int (p_\mathrm{r} - p_\mathrm{avg}).
\end{align}

\subsubsection{Grid-side current controller} It consists of two PI controllers with PI gains $k^p_\mathrm{GCC}, k^i_\mathrm{GCC}$, and they output the references for the inverter terminal voltage:
 \begin{align}
v_\mathrm{i}^\mathrm{d*} &= k^p_\mathrm{GCC} (i_\mathrm{i}^\mathrm{d*} - i_\mathrm{i}^\mathrm{d}) + k^i_\mathrm{GCC} \int (i_\mathrm{i}^\mathrm{d*} - i_\mathrm{i}^\mathrm{d}), \\
v_\mathrm{i}^\mathrm{q*} &= k^p_\mathrm{GCC} (i_\mathrm{i}^\mathrm{q*} - i_\mathrm{i}^\mathrm{q}) + k^i_\mathrm{GCC} \int (i_\mathrm{i}^\mathrm{q*} - i_\mathrm{i}^\mathrm{q}).
\end{align}
For an ideal inverter, $v_\mathrm{i}^\mathrm{d*} = v_\mathrm{i}^\mathrm{d}$, $v_\mathrm{i}^\mathrm{q*} = v_\mathrm{i}^\mathrm{q}$.

\subsection{DC-link capacitor} The DC-link capacitor filters out the variations of the DC voltage (output of the rectifier) prior to further processing by the inverter. This is sketched in Fig.~\ref{fig:blockdiagram} as the capacitor in between the switching blocks labeled RSC and the GSC. If we consider the energy stored at the capacitor, denoted by $E_\mathrm{C}$, as a state, then the dynamics of the capacitor are
\begin{equation}
\dot{E}_\mathrm{C} =  \frac{1}{C} (p_\mathrm{r} - p_\mathrm{avg}).
\end{equation}
Here, we use the filtered version of the grid-injected power instead of the power corresponding to the inverter terminal voltage $v_\mathrm{i}$ since the $LCL$ filter is lossless. Notice that the power controller in the grid-side converter also regulates the DC-link capacitor voltage such that its energy stored, $E_\mathrm{C}$ reaches its steady-state value asymptotically. 

\subsection{State-space Model for the Wind Turbine System}
For notational and expositional convenience, we define:
\begin{align}
&\dot{\phi}_\mathrm{r}^\mathrm{q} = q_\mathrm{s}^* - q_\mathrm{s},\, \dot{\phi}_\mathrm{r}^\mathrm{t} = T_\mathrm{e}^* - T_\mathrm{e},\, \dot{\phi}_\mathrm{r}^\mathrm{id} = i_\mathrm{r}^\mathrm{d*} - i_\mathrm{r}^\mathrm{d}, \\
&\dot{\phi}_\mathrm{r}^\mathrm{iq} = i_\mathrm{r}^\mathrm{q*} - i_\mathrm{r}^\mathrm{q},\, \phi_\mathrm{g}^\mathrm{p} = P_\mathrm{r} - p_\mathrm{avg},\, \dot{\phi}_\mathrm{g}^\mathrm{q} = q^* - q_\mathrm{i}, \\
&\dot{\phi}_\mathrm{g}^\mathrm{id} = i_\mathrm{i}^\mathrm{d*} - i_\mathrm{i}^\mathrm{d},\, \dot{\phi}_\mathrm{g}^\mathrm{iq} = i_\mathrm{i}^\mathrm{q*} -  i_\mathrm{i}^\mathrm{q}.
\end{align}
Then, the dynamics of an individual wind turbine can be compactly represented in the following state-space form:
\begin{equation}
\dot{x} = A x + B u_1 + g(x,u_2,u_3),
\label{eq:nominal}
\end{equation}
where the states and inputs are given by
\begin{align*}
\label{eq:IndividualModel}
x &= [i_\mathrm{s}^\mathrm{d}, i_\mathrm{s}^\mathrm{q}, \phi_\mathrm{r}^\mathrm{q}, \phi_\mathrm{r}^\mathrm{t}, \phi_\mathrm{r}^\mathrm{id}, \phi_\mathrm{r}^\mathrm{iq}, i_\mathrm{i}^\mathrm{d}, i_\mathrm{i}^\mathrm{q}, i_\mathrm{g}^\mathrm{d}, i_\mathrm{g}^\mathrm{q}, \phi_\mathrm{g}^\mathrm{id}, \phi_\mathrm{g}^\mathrm{iq}, p_\mathrm{avg}, q_\mathrm{avg}, \\
&\quad  \phi_\mathrm{g}^\mathrm{p}, \phi_\mathrm{g}^\mathrm{q}, \omega_\mathrm{r}, \omega_\mathrm{t}, \theta_\mathrm{tw}, e_\mathrm{s}^\mathrm{d}, e_\mathrm{s}^\mathrm{q}, v_{\mathrm{f}}^{\mathrm{d}},  v_{\mathrm{f}}^{\mathrm{q}}, v_{\mathrm{PLL}}, \phi_{\mathrm{PPL}}, \delta,  E_\mathrm{C}]^\mathrm{T},
\end{align*}
\begin{equation*}
u_1 = q^*,\quad u_2 = [v_\mathrm{g}^\mathrm{a}, v_\mathrm{g}^\mathrm{b}, v_\mathrm{g}^\mathrm{c}]^\mathrm{T},\quad 
u_3 = v_{\mathrm{w}}.
\end{equation*}
The entries of the matrices $A \in \mathbb{R}^{27 \times 27}$, $B \in \mathbb{R}^{27 }$, and function $g: \mathbb{R}^{27} \times \mathbb{R}^{3}\times \mathbb{R} \to \mathbb{R}^{27}$ follow from the models discussed in Section~\ref{sec:Model}, and they are listed in Appendix~A. 

\section{Reduced-order Aggregate Model}
\label{sec: ROM}
We begin this section by introducing scalings for a subset of the control parameters, filter elements, and constants of the individual wind turbine model which yield the aggregate model for the parallel system. Following this, we present the main result of the paper.  

\subsection{Parametric Scalings and State-space Model of the Reduced-order Model}
We consider $N$ identical wind turbines with DFIGs, rotor-side, and grid-side converters connected in parallel to the grid as shown in Fig.~\ref{fig:parallel}. We assume that all wind turbines have the same incident wind speed, $v_\mathrm w$, and the reactive-power setpoints for the rotor side converters, $q^* = 0$. Consider the following parametric scalings:
\begin{align}
\begin{split}
&\left( H_\mathrm{t}, H_\mathrm{g}, c_\mathrm{sh}, k_\mathrm{sh}, T_\mathrm{m,base}, R_1, R_2, L_\mathrm{s}', X_\mathrm{m}, k^{p\mathrm{d}}_\mathrm{RCC},  \right.\\ 
& \left.  k^{i\mathrm{d}}_\mathrm{RCC}, k^{p\mathrm{q}}_\mathrm{RCC}, k^{i\mathrm{q}}_\mathrm{RCC}, K_\mathrm{opt}, L_\mathrm{i},  C_\mathrm{f}, L_\mathrm{g}, k^p_\mathrm{GCC},k^i_\mathrm{GCC},  \right. \\
& \left. C  \right) \to  \left( N H_\mathrm{t}, N H_\mathrm{g}, N c_\mathrm{sh}, N k_\mathrm{sh}, \frac{T_\mathrm{m,base}}{N}, \frac{R_1}{N}, \frac{R_2}{N},   \right.  \\
&\left.  \frac{L_\mathrm{s}'}{N},  \frac{X_\mathrm{m}}{N},  \frac{k^{p\mathrm{d}}_\mathrm{RCC}}{N}, \frac{k^{i\mathrm{d}}_\mathrm{RCC}}{N}, \frac{k^{p\mathrm{q}}_\mathrm{RCC}}{N}, \frac{k^{i\mathrm{q}}_\mathrm{RCC}}{N}, N K_\mathrm{opt}, \frac{L_\mathrm{i}}{N},   \right. \\
&\left.   N C_\mathrm{f}, \frac{L_\mathrm{g}}{N}, \frac{k^p_\mathrm{GCC}}{N}, \frac{k^i_\mathrm{GCC}}{N}, NC \right).
\end{split} \label{eq:scalings}
\end{align}
Then, the mechanical and electrical dynamics of the $N$-turbine collection can be captured by:
\begin{equation}
\dot{x}^\mathrm{r} = A^\mathrm{r} x^\mathrm{r} + B^\mathrm{r} u_1^\mathrm{r} + g^\mathrm{r}(x^\mathrm{r},u_2^\mathrm{r},u_3^\mathrm{r}),
\label{eq:aggregate}
\end{equation}
where, the states and the inputs are:
\begin{align*}
x^\mathrm{r} &= [i_\mathrm{s}^\mathrm{d,r}, i_\mathrm{s}^\mathrm{q,r}, \phi_\mathrm{r}^\mathrm{q,r}, \phi_\mathrm{r}^\mathrm{t,r}, \phi_\mathrm{r}^\mathrm{id,r}, \phi_\mathrm{r}^\mathrm{iq,r}, i_\mathrm{i}^\mathrm{d,r}, i_\mathrm{i}^\mathrm{q,r}, i_\mathrm{g}^\mathrm{d,r}, i_\mathrm{g}^\mathrm{q,r}, \phi_\mathrm{g}^\mathrm{id,r},  \\
&\quad  \phi_\mathrm{g}^\mathrm{iq,r}, p_\mathrm{avg}^\mathrm r, q_\mathrm{avg}^\mathrm r, \phi_\mathrm{g}^\mathrm{p,r}, \phi_\mathrm{g}^\mathrm{q,r}, \omega_\mathrm{r}^\mathrm{r}, \omega_\mathrm{t}^\mathrm{r}, \theta_\mathrm{tw}^\mathrm{r}, e_\mathrm{s}^\mathrm{d,r}, e_\mathrm{s}^\mathrm{q,r}, v_{\mathrm{f}}^{\mathrm{d,r}}, \\
&\quad v_{\mathrm{f}}^{\mathrm{q,r}}, v_{\mathrm{PLL}}^\mathrm{r}, \phi_{\mathrm{PPL}}^\mathrm{r}, \delta^\mathrm{r},  E_\mathrm{C}^\mathrm{r}]^\mathrm{T}, \\
u_1^\mathrm{r} &= N u_1 = Nq^*,\, u_2^\mathrm{r} = u_2 = [v_\mathrm{g}^\mathrm{a}, v_\mathrm{g}^\mathrm{b}, v_\mathrm{g}^\mathrm{c}]^\mathrm{T},\, u_3^\mathrm{r} = u_3 = v_{\mathrm{w}},
\end{align*}
and the matrices $A^\mathrm{r} \in \mathbb{R}^{27 \times 27}$, $B^\mathrm{r} \in \mathbb{R}^{27}$, and function $g^\mathrm{r}: \mathbb{R}^{27} \times \mathbb{R}^3 \times \mathbb{R} \to \mathbb{R}^{27}$ have the same structure as $A$, $B$, and $g$ for the individual model as given in~\eqref{eq:nominal}.

Our central claim is that the state-space model in~\eqref{eq:aggregate} obtained with the parametric scalings in~\eqref{eq:scalings} generates dynamics that are consistent with the electrical interconnection shown in Fig.~\ref{fig:parallel} and the assumption that all turbines see the same incident wind. In particular, currents and power-related states in the aggregate model are $N$ times those in the individual turbine, while voltages, PLL-related, and turbine-related states are the same. (This is illustrated for a subset of states in Fig.~\ref{fig:blockdiagram}.) We formally establish this next.

\subsection{Validating the Reduced-order Model}
Here, we establish the relationship between the states of the aggregated model in~\eqref{eq:aggregate} and the model for individual turbines in~\eqref{eq:nominal}. To do so, we will find the following definition of a \emph{scaling vector} $\psi := [N 1_{16}^\mathrm{T}, 1_{11}^\mathrm{T}]^\mathrm{T}$ useful.

\begin{theorem} (Aggregate Model Validation.) Consider the dynamics of the individual wind-turbine model in~\eqref{eq:nominal} and the corresponding reduced-order model in~\eqref{eq:aggregate} with the same structure and dimension. \emph{If}: i)~the reduced-order model in~\eqref{eq:aggregate} is synthesized with the parametric scalings given in~\eqref{eq:scalings}, ii)~the initial conditions of the states are chosen such that $ x^\mathrm{r}(t_0) = \mathrm{diag}(\psi) x(t_0)$, and iii)~the inputs are such that $u_1^\mathrm{r} = N u_1$, $u_2^\mathrm{r} = u_2,$ $u_3^\mathrm{r} = u_3$, \emph{then} it follows that $ x^\mathrm{r}(t) = \mathrm{diag}(\psi)x(t)$, $\forall t \ge t_0$.

\end{theorem}

Given the definition of the scaling vector, $\psi$, the above statement implies that all the currents, powers, and associated controller-related states in the aggregated model are $N$ times that in the individual wind-turbine model: 
\begin{align*} 
&[i_\mathrm{s}^\mathrm{d,r}, i_\mathrm{s}^\mathrm{q,r}, \phi_\mathrm{r}^\mathrm{q,r}, \phi_\mathrm{r}^\mathrm{t,r}, \phi_\mathrm{r}^\mathrm{id,r}, \phi_\mathrm{r}^\mathrm{iq,r}, i_\mathrm{i}^\mathrm{d,r}, i_\mathrm{i}^\mathrm{q,r}, i_\mathrm{g}^\mathrm{d,r}, i_\mathrm{g}^\mathrm{q,r}, \phi_\mathrm{g}^\mathrm{id,r},  \\
&  \phi_\mathrm{g}^\mathrm{iq,r}, p_\mathrm{avg}^\mathrm r, q_\mathrm{avg}^\mathrm r, \phi_\mathrm{g}^\mathrm{p,r}, \phi_\mathrm{g}^\mathrm{q,r}]^\mathrm{T} := N [i_\mathrm{s}^\mathrm{d}, i_\mathrm{s}^\mathrm{q}, \phi_\mathrm{r}^\mathrm{q}, \phi_\mathrm{r}^\mathrm{t}, \phi_\mathrm{r}^\mathrm{id}, \phi_\mathrm{r}^\mathrm{iq}, i_\mathrm{i}^\mathrm{d},\\
& i_\mathrm{i}^\mathrm{q}, i_\mathrm{g}^\mathrm{d}, i_\mathrm{g}^\mathrm{q}, \phi_\mathrm{g}^\mathrm{id}, \phi_\mathrm{g}^\mathrm{iq}, p_\mathrm{avg}, q_\mathrm{avg}, \phi_\mathrm{g}^\mathrm{p}, \phi_\mathrm{g}^\mathrm{q}]^\mathrm{T}.
\end{align*}
Furthermore, turbine-related states, voltages, and PLL-related states in the aggregated and individual-turbine models evolve identically: 
\begin{align*} 
&[\omega_\mathrm{r}^\mathrm{r}, \omega_\mathrm{t}^\mathrm{r}, \theta_\mathrm{tw}^\mathrm{r}, e_\mathrm{s}^\mathrm{d,r}, e_\mathrm{s}^\mathrm{q,r}, v_{\mathrm{f}}^{\mathrm{d,r}}, 
v_{\mathrm{f}}^{\mathrm{q,r}}, v_{\mathrm{PLL}}^\mathrm{r}, \phi_{\mathrm{PPL}}^\mathrm{r}, \delta^\mathrm{r},  E_\mathrm{C}^\mathrm{r}]^\mathrm{T}: =\\
&[ \omega_\mathrm{r}, \omega_\mathrm{t}, \theta_\mathrm{tw}, e_\mathrm{s}^\mathrm{d}, e_\mathrm{s}^\mathrm{q}, v_{\mathrm{f}}^{\mathrm{d}},  v_{\mathrm{f}}^{\mathrm{q}}, v_{\mathrm{PLL}}, \phi_{\mathrm{PPL}}, \delta,  E_\mathrm{C}]^\mathrm{T}.
\end{align*} 
This is consistent with the parallel electrical interconnection (Fig.~\ref{fig:parallel}) and the assumption that all turbines see the same incident wind and the associated rotor-side converters have the same reactive-power setpoints.
  
\begin{proof}
We first partition $x = [x_{\mathrm{T}}^\mathrm{T},x_{\mathrm{D}}^\mathrm{T}, x_\mathrm{I}^\mathrm{T}, E_\mathrm{C}]^\mathrm{T}$, where
$x_\mathrm{T} := [\omega_\mathrm{r}, \omega_\mathrm{t}, \theta_\mathrm{tw}]^\mathrm{T}$, $x_\mathrm{D} := [i_\mathrm{s}^\mathrm{d}, i_\mathrm{s}^\mathrm{q}, \phi_\mathrm{r}^\mathrm{q}, \phi_\mathrm{r}^\mathrm{t}, \phi_\mathrm{r}^\mathrm{id}, \phi_\mathrm{r}^\mathrm{iq}, e_\mathrm{s}^\mathrm{d}, e_\mathrm{s}^\mathrm{q}]^\mathrm{T}$, $x_\mathrm{I} := [i_\mathrm{i}^\mathrm{d}, i_\mathrm{i}^\mathrm{q}, i_\mathrm{g}^\mathrm{d}, i_\mathrm{g}^\mathrm{q}, \phi_\mathrm{g}^\mathrm{id}, \phi_\mathrm{g}^\mathrm{iq}, p_\mathrm{avg}, q_\mathrm{avg},\phi_\mathrm{g}^\mathrm{p}, \phi_\mathrm{g}^\mathrm{q},v_{\mathrm{f}}^{\mathrm{d}},  v_{\mathrm{f}}^{\mathrm{q}}, v_{\mathrm{PLL}},\\ \phi_{\mathrm{PPL}},$ $ \delta]^\mathrm{T}$, capture the turbine, DFIG \& rotor-side converter, and grid-side-converter states; and we also partition $x^\mathrm{r}$ the same way. We next investigate these dynamics sequentially.  

\subsubsection*{Turbine} First, consider the turbine aerodynamic model in~\eqref{eq:aerodynamics1}--\eqref{eq:aerodynamics3}. These can be expressed in state-space form:
\begin{equation}
\dot{x}_\mathrm{T} = A_\mathrm{T} x_\mathrm{T} + g_\mathrm{T}(x,u_2,u_3),
\end{equation}
where $A_{\mathrm T}$ and $g_{\mathrm T}$  are constructed from appropriate entries of $A$ and $g$ spelled out in Appendix~A. Define $z_\mathrm{T} := x_\mathrm{T}^\mathrm{r} -  x_\mathrm{T}$. Then, we get
\begin{align}
\dot{z}_\mathrm{T} &= \dot{x}_\mathrm{T}^\mathrm{r} - \dot{x}_\mathrm{T} = A^\mathrm{r}_\mathrm{T} x^\mathrm{r} + g_\mathrm{T}^\mathrm{r}(x^\mathrm{r},u_2^\mathrm{r},u_3^\mathrm{r}) - A_\mathrm{T} x_\mathrm{T}  \nonumber \\
& - g_\mathrm{T}(x,u_2,u_3). 
\label{eq:z_T}
\end{align}
With the parametric scalings spelled out in~\eqref{eq:scalings}, we have $A_\mathrm{T} = A_\mathrm{T}^\mathrm{r}$. Furthermore, for the nonlinear elements in the model:
\begin{align}
g_{\mathrm{T},1}^\mathrm{r} &= \frac{-1}{2NH_\mathrm{g}}\bigg( \frac{v_\mathrm{g} ^ \mathrm{q}}{\omega_\mathrm{s}}Ni_\mathrm{s}^\mathrm{q} + \frac{v_\mathrm{g}^\mathrm{d}}{\omega_\mathrm s}Ni_\mathrm{s}^\mathrm{d} \bigg) = g_{\mathrm{T},1}, \label{eq:g1}
 \\ 
g_{\mathrm{T},2}^\mathrm{r} &= 0 =g_{\mathrm{T},2},\label{eq:g2} \\
g_{\mathrm{T},3}^\mathrm r &= \frac{N}{4T_{\mathrm{m,base}}NH_\mathrm{t}}\left(\rho \pi R^2 C_p(\lambda,\beta)\frac{v_{\mathrm{w}}^3}{\omega_\mathrm{t}}\right) = g_{\mathrm{T},3}. \label{eq:g3}
\end{align}
While~\eqref{eq:g2} is obvious by definition,~\eqref{eq:g1} follows from the scaling adopted for $H_\mathrm g$ in \eqref{eq:scalings}, and the fact that currents $i_\mathrm{s}^\mathrm{d,r}$ and $i_\mathrm{s}^\mathrm{q,r}$ are $N$ times the individual model because the turbines are connected in parallel. Finally,~\eqref{eq:g3} follows from the scaling adopted for $T_{\mathrm{m,base}}$ and the fact that incident wind speeds are all the same. If the initial conditions, $z_\mathrm{T}(t_0) = 0_3$, then $z_\mathrm{T} = 0_3$, $\forall t\geq 0$. This precisely implies that $x_\mathrm{T} = x_\mathrm{T}^\mathrm{r}$, or in other words, dynamics of the turbine for the aggregated and individual turbines are identical.

\subsubsection*{DFIG \& Rotor-side Converter} We now proceed with the next part of the proof that focuses on the dynamics of the DFIG and rotor-side converters. As before, define $z_\mathrm{D} := x_\mathrm{D}^\mathrm{r} - \mathrm{diag}(\psi_\mathrm{D}) x_\mathrm{D}$, where $\psi_\mathrm{D} := [N 1_6^\mathrm{T}, 1_2^\mathrm{T}]^\mathrm{T}$. The dynamics of $z_\mathrm{D}$ are given by
\begin{align}
\dot{z}_\mathrm{D} &= \dot{x}_\mathrm{D} - \mathrm{diag}(\psi_\mathrm{D}) \dot{x}_\mathrm{D} = A^\mathrm{r}_\mathrm{D} x^\mathrm{r} + B^\mathrm{r}_{\mathrm{D}} u_1^\mathrm{r} \nonumber \\
&+ g_\mathrm{D}^\mathrm{r}(x_\mathrm{D}^\mathrm{r},u_2^\mathrm{r},u_3^\mathrm{r}) - \mathrm{diag}(\psi_\mathrm{D})A_\mathrm{D} - \mathrm{diag}(\psi_\mathrm{D}) B_{\mathrm{D}} u_1^\mathrm{r} \nonumber \\
& - \mathrm{diag}(\psi_\mathrm{D})g_\mathrm{D}(x_\mathrm{D},u_2,u_3),
\label{eq:z_D}
\end{align}
where $A_\mathrm{D} \in \mathbb{R}^{8 \times 8}$, $B_{\mathrm{D}} \in \mathbb{R}^8$, and $g_\mathrm{D}: \mathbb{R}^{8} \times \mathbb{R}^{3} \times \mathbb{R} \to \mathbb{R}^{8}$ (and similarly  $A_\mathrm{D}^\mathrm{r}$, $B_{\mathrm{D}}^\mathrm{r}$, and $g_\mathrm{D}^\mathrm{r}$) are derived from appropriate entries of \eqref{eq:nominal}. Furthermore, from Section~\ref{subsec:generator} and Section~\ref{subsec:rotor-converter} we also note that the dynamics of the DFIG and rotor-side converters are decoupled from the remainder of the states. We will show that $\dot{z}_\mathrm{D} = 0_8, \forall t \ge t_0$ when $z_\mathrm{D}(t_0) =  x_\mathrm{D}^\mathrm{r}(t_0) - \mathrm{diag}(\psi_\mathrm{D}) x_\mathrm{D}(t_0) = 0_{8}$, which implies $x_\mathrm{D}^\mathrm{r}(t) = \mathrm{diag}(\psi_\mathrm{D}) x_\mathrm{D}(t), \forall t \ge t_0$. To this end, partition $x_\mathrm{D} = [x_{\mathrm{D},1}^\mathrm{T}, x_{\mathrm{D},2}^\mathrm{T}]^\mathrm{T}$, where $x_{\mathrm{D},1} := [i_\mathrm{s}^\mathrm{d}, i_\mathrm{s}^\mathrm{q}, \phi_\mathrm{r}^\mathrm{q}, \phi_\mathrm{r}^t, \phi_\mathrm{r}^\mathrm{id}, \phi_\mathrm{r}^\mathrm{iq}]^\mathrm{T}$ and $x_{\mathrm D,2} := [ e_\mathrm{s}^\mathrm{d}, e_\mathrm{s}^\mathrm{q}]^\mathrm{T}$, and similarly for $x_\mathrm{D}^\mathrm{r}$. Then, the dynamics of $x_\mathrm{D}$ and $x_\mathrm{D}^\mathrm{r}$ are partitioned as follows:
\begin{align*}
\begin{bmatrix} \dot{x}_{\mathrm{D},1}\\ \dot{x}_{\mathrm{D},2} \end{bmatrix}
&=
\begin{bmatrix} A_{11} & A_{12}\\ A_{21} & A_{22} \end{bmatrix}
\begin{bmatrix} x_{\mathrm{D},1}\\ x_{\mathrm{D},2} \end{bmatrix}
+
\begin{bmatrix} B_{1}\\ B_{2} \end{bmatrix} u_1
+ g_\mathrm{D}(x_\mathrm{D},u_2,u_3), \\
\begin{bmatrix} \dot{x}_{\mathrm{D},1}^\mathrm{r} \\ \dot{x}_{\mathrm{D},2}^\mathrm{r} \end{bmatrix}
&=
\begin{bmatrix} A_{11}^\mathrm{r} & A_{12}^\mathrm{r}\\ A_{21}^\mathrm{r} & A_{22}^\mathrm{r} \end{bmatrix}
\begin{bmatrix} x_{\mathrm{D},1}^\mathrm{r}\\ x_{\mathrm{D},2}^\mathrm{r} \end{bmatrix}
+
\begin{bmatrix} B_{1}^\mathrm{r}\\ B_{2}^\mathrm{r} \end{bmatrix} u_1^\mathrm{r}
+ g_\mathrm{D}^\mathrm{r}(x^\mathrm{r}_\mathrm{D},u_2^\mathrm{r},u_3^\mathrm{r}). 
\end{align*}
From the definition of the state-space matrices in the appendix and the parametric scalings in~\eqref{eq:scalings}, we get:
\begin{align*}
&A_{11}^\mathrm{r} = A_{11},\quad A_{12}^\mathrm{r} = N A_{12},\quad A_{21}^\mathrm{r} = \frac{1}{N} A_{21},\\
& A_{22}^\mathrm{r} = A_{22},\quad B_{1}^\mathrm{r} = B_{1},\quad B_{2}^\mathrm{r} = \frac{1}{N} B_{2}.
\end{align*}
Then, we have 
\begin{align}
\mathrm{diag}(\psi_\mathrm{D}) A_\mathrm{D} &= 
\begin{bmatrix} N A_{11} & N A_{12}\\ A_{21} &  A_{22} \end{bmatrix} 
= 
\begin{bmatrix}
N A_{11}^\mathrm{r} &  A_{12}^\mathrm{r}\\ N A_{21}^\mathrm{r} &  A_{22}^\mathrm{r}
\end{bmatrix}\nonumber \\
&= 
A_\mathrm{D}^\mathrm{r} \mathrm{diag}(\psi_\mathrm{D}), \label{eq:identity1} \\
\mathrm{diag}(\psi_D) B &= 
\begin{bmatrix} N B_{1} \\ B_{2} \end{bmatrix} 
= 
\begin{bmatrix} N B_{1}^\mathrm{r} \\ N B_{2}^\mathrm{r} \end{bmatrix}
= N B^\mathrm{r}. \label{eq:identity2}
\end{align}
A similar relationship is obtained for the nonlinearities:
\begin{align}
\mathrm{diag}(\psi_\mathrm{D})g_\mathrm{D}(x_\mathrm{D},u_2,u_3) = g^\mathrm{r}_\mathrm{D}(\mathrm{diag}(\psi_\mathrm{D})x_\mathrm{D},u_2^\mathrm{r}, u_3^\mathrm{r}).
\label{eq:identity4}
\end{align}
Using the identities~\eqref{eq:identity1}--\eqref{eq:identity4} in~\eqref{eq:z_D}, we have
\begin{align}
\dot{z}_\mathrm{D} &= A_\mathrm{D}^\mathrm{r} z_\mathrm{D} + g^\mathrm{r}_\mathrm{D}(x^\mathrm{r}_\mathrm{D},u^\mathrm{r}_2,u^\mathrm{r}_3) - g_\mathrm{D}^\mathrm{r}(\mathrm{diag}(\psi_\mathrm{D})x_\mathrm{D},u_2^\mathrm{r},u_3^\mathrm{r}) \nonumber \\
&=  A_\mathrm{D}^\mathrm{r} z_\mathrm{D} + f(z_\mathrm{D},u_2^\mathrm{r},u_3^\mathrm{r}),
\end{align}
where $f(0_{8},u_2^\mathrm{r},u_3^\mathrm{r}) = 0_{8}$. If $z_\mathrm{D}(t_0) = 0_{8}$, then $z_\mathrm{D}(t) = x_\mathrm{D}^\mathrm{r}(t) - \mathrm{diag}(\psi_\mathrm{D}) x_\mathrm{D}(t) = 0_{8}, \forall t \ge t_0$.
\subsubsection*{Grid-side converter} For the inverter states (that include those of the PLL) $x_\mathrm{I}$, we have proved in~\cite{Purba-Allerton-2017} that $x_\mathrm{I}^\mathrm{r}(t) = \mathrm{diag}(\psi_\mathrm{I}) x_\mathrm{I}(t), \forall t \ge t_0$, where $\psi_\mathrm{I} = [N 1_{10}^\mathrm{T}, 1_5^\mathrm{T}]$, when $x_\mathrm{I}^\mathrm{r}(t_0) = \mathrm{diag}(\psi_\mathrm{I}) x_\mathrm{I}(t_0)$, the reference power setpoints of the aggregated inverter are $N$ times that of the individual, and the inverters are all in parallel which means they sense the same grid voltage. For the wind-turbine setting, the reference active power to the inverter is  not a constant setpoint, but it is set to the rotor active power $p_\mathrm{r}$. Recall that the states of the rotor-side converter are decoupled from the grid-side converter. Therefore, if we show that $p^\mathrm{r}_\mathrm{r}(t) = N p_\mathrm{r}(t), \forall t \ge t_0$, where $p^\mathrm{r}_\mathrm{r}$ denotes the rotor active power of the aggregate model, then the analysis in~\cite{Purba-Allerton-2017} still holds and $x_\mathrm{I}^\mathrm{r}(t) = \mathrm{diag}(\psi_\mathrm{I}) x_\mathrm{I}(t), \forall t \ge t_0$. The rotor active power $p_\mathrm{r}$ is given in~\eqref{eq:rotor_power}, with $v_\mathrm{r}^\mathrm{d}$, $v_\mathrm{r}^\mathrm{q}$, $i_\mathrm{r}^\mathrm{d}$, and $i_\mathrm{r}^\mathrm{q}$ listed in~\eqref{eq:vr_d}--\eqref{eq:ir}. The complete form of $v_\mathrm{r}^\mathrm{d}$ and $v_\mathrm{r}^\mathrm{q}$ (i.e., in terms of the states and inputs) are:
\begin{align}
v_\mathrm{r}^\mathrm{d} &= k^{p\mathrm{d}}_\mathrm{RCC}(k^p_\mathrm{RPC}(q^*_\mathrm{s} - (-v_\mathrm{s}^\mathrm{q} i_\mathrm{s}^\mathrm{d} + v_\mathrm{s}^\mathrm{d} i_\mathrm{s}^\mathrm{q})) + k^i_\mathrm{RPC} \phi_\mathrm{r}^\mathrm{q} - i_\mathrm{r}^\mathrm{d}) \nonumber \\
&\quad + k^{i\mathrm{q}}_\mathrm{RCC} \phi_\mathrm{r}^\mathrm{id}, \\
v_\mathrm{r}^\mathrm{q} &= k^{p\mathrm{q}}_\mathrm{RCC}(k^p_\mathrm{RPC}(K_\mathrm{opt}\omega_\mathrm{r}^2 - (\frac{e_\mathrm{s}^\mathrm{d}}{\omega_\mathrm{s}} i_\mathrm{s}^\mathrm{d} + \frac{e_\mathrm{s}^\mathrm{q}}{\omega_\mathrm{s}} i_\mathrm{s}^\mathrm{q})) + k^i_\mathrm{RPC} \phi_\mathrm{r}^\mathrm{t} \nonumber \\
&\quad - i_\mathrm{r}^\mathrm{q}) + k^{i\mathrm{q}}_\mathrm{RCC} \phi_\mathrm{r}^\mathrm{iq} .
\end{align}
Then, the ($\mathrm{dq}$) rotor voltage and current in the reduced-order model, denoted by $v_\mathrm{r}^\mathrm{d,r}$, $v_\mathrm{r}^\mathrm{q,r}$, $i_\mathrm{r}^\mathrm{d,r}$, and $i_\mathrm{r}^\mathrm{d,r}$, are given by
\begin{align}
v_\mathrm{r}^\mathrm{d,r} &= \frac{ k^{p\mathrm d}_\mathrm{RCC}}{N}(k^{p}_\mathrm{RPC}(N q^*_\mathrm{s} - (-v_\mathrm{s}^\mathrm{q} i_\mathrm{s}^\mathrm{d,r} + v_\mathrm{s}^\mathrm{d} i_\mathrm{s}^\mathrm{q,r}))   \nonumber \\
&\quad + k^i_\mathrm{RPC} \phi_\mathrm{r}^\mathrm{q,r} - i_\mathrm{r}^\mathrm{d,r}) + \frac{k^{i\mathrm d}_\mathrm{RCC}}{N}  \phi_\mathrm{r}^\mathrm{id,r}, \label{eq:vr_dr} \\
v_\mathrm{r}^\mathrm{q,r} &= \frac{k^{p\mathrm q}_\mathrm{RCC}}{N} (k^p_\mathrm{RPC}(N K_\mathrm{opt}(\omega_\mathrm{r}^\mathrm{r})^2 - (\frac{e_\mathrm{s}^\mathrm{d,r}}{\omega_\mathrm{s}} i_\mathrm{s}^\mathrm{d,r} + \frac{e_\mathrm{s}^\mathrm{q,r}}{\omega_\mathrm{s}} i_\mathrm{s}^\mathrm{q,r}))   \nonumber \\
&\quad + k^i_\mathrm{RPC} \phi_\mathrm{r}^\mathrm{t,r} - i_\mathrm{r}^\mathrm{q,r}) + \frac{k^{i\mathrm q}_\mathrm{RCC}}{N} \phi_\mathrm{r}^\mathrm{iq,r}, \label{eq:vr_qr}  \\
i_\mathrm{r}^\mathrm{d,r} &= \frac{N e_\mathrm{s}^\mathrm{q,r}}{X_\mathrm{m}} - K_\mathrm{mrr}i_\mathrm{s}^\mathrm{d,r},\, i_\mathrm{r}^\mathrm{q,r} = \frac{N e_\mathrm{s}^\mathrm{d,r}}{X_\mathrm{m}} - K_\mathrm{mrr}i_\mathrm{s}^\mathrm{q,r} \label{eq:ir_r} .
\end{align}
From the proof of the turbine and DFIG \& rotor-side converter, we have the following relationships:
\begin{align*}
&i_\mathrm{s}^\mathrm{d,r} = N i_\mathrm{s}^\mathrm{d},\, i_\mathrm{s}^\mathrm{q,r} = N i_\mathrm{s}^\mathrm{q},\, \phi_\mathrm{r}^\mathrm{q,r} = N\phi_\mathrm{r}^\mathrm{q},\, \phi_\mathrm{r}^\mathrm{t,r} = N \phi_\mathrm{r}^\mathrm{t},\, \\
&\phi_\mathrm{r}^\mathrm{id,r} = N \phi_\mathrm{r}^\mathrm{id,r},
 \phi_\mathrm{r}^\mathrm{iq,r} = N \phi_\mathrm{r}^\mathrm{iq},\, e_\mathrm{s}^\mathrm{d,r} =  e_\mathrm{s}^\mathrm{d},\,e_\mathrm{s}^\mathrm{q,r} =  e_\mathrm{s}^\mathrm{q},\,\omega_\mathrm{r}^\mathrm{r} = \omega_\mathrm{r}.
\end{align*}
Substituting these into~\eqref{eq:vr_dr}--\eqref{eq:ir_r}, it is straightforward to show the following $\forall t \ge t_0$:
\begin{align*}
v_\mathrm{r}^\mathrm{d,r}(t) &= v_\mathrm{r}^\mathrm{d}(t),\quad v_\mathrm{r}^\mathrm{q,r}(t) = v_\mathrm{r}^\mathrm{q}(t), \\
i_\mathrm{r}^\mathrm{d,r}(t) &= N i_\mathrm{r}^\mathrm{d}(t),\quad i_\mathrm{r}^\mathrm{q,r}(t) = N i_\mathrm{r}^\mathrm{q}(t).
\end{align*}
Then, we have $p_\mathrm{r}^\mathrm{r}(t) =  N p_\mathrm{r}(t), \forall t \ge t_0$. 

\subsubsection*{DC-link Capacitance} Lastly, we define $z_\mathrm{C} := E_\mathrm{C}^\mathrm{r} - E_\mathrm{C}$. Then, $z_\mathrm{C}$ evolves as below:
\begin{equation}
\dot{z}_\mathrm{C} = \dot{E}_\mathrm{C}^\mathrm{r} - \dot{E}_\mathrm{C} = \frac{1}{NC}(p_\mathrm{r}^\mathrm{r} - p_\mathrm{avg}^\mathrm{r}) - \frac{1}{C}(p_\mathrm{r} - p_\mathrm{avg}).
\end{equation}
We note that the dynamics of the rotor-side and grid-side converters are decoupled from the DC-link capacitor. In the previous section focusing on the grid-side converter, we have proved that $p^\mathrm{r}_\mathrm{r}(t) = N p_\mathrm{r}(t), p_\mathrm{avg}^\mathrm{r}(t) = N p_\mathrm{avg}(t), \forall t \ge t_0$. Then if we initialize $z_\mathrm{C}(t_0) = E_\mathrm{C}^\mathrm{r}(t_0) - E_\mathrm{C}(t_0) = 0$, we have $z_\mathrm{C}(t) = E_\mathrm{C}^\mathrm{r}(t) - E_\mathrm{C}(t) = 0, \forall t \ge t_0$. 

From the above conclusions for the dynamics of the turbine, DFIG \& rotor-side converters, grid-side converters, and DC-link capacitance, we can surmise that $ x^\mathrm{r}(t) = \mathrm{diag}(\psi)x(t)$, $\forall t \ge t_0$.
\end{proof}

\begin{figure}
\centering
\includegraphics[scale=0.4] {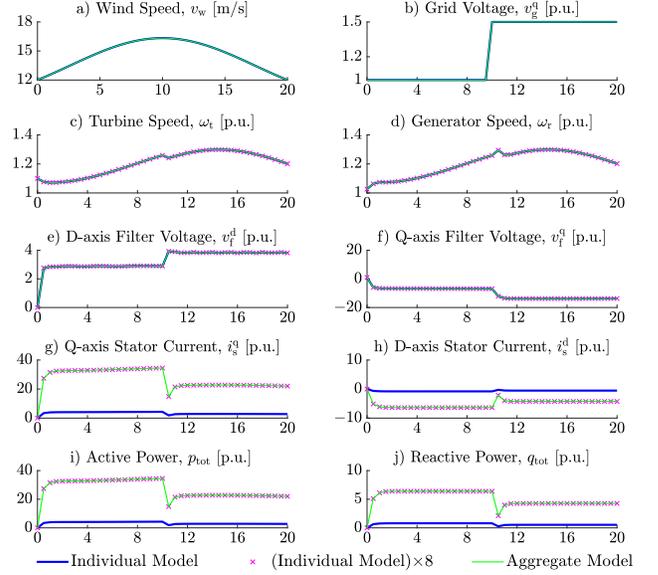}
\caption{Simulation results for the aggregated model and multiple instances of the original model. For the wind speed and grid voltage shown in (a) and (b): mechanical variables and voltages (c)-(f) are the same for the original and reduced-order models, while currents and powers are scaled (g)-(j).}
\label{fig:plots}
\end{figure}


\section{Numerical Simulation Results}
\label{sec: Results}
In this section, we present numerical simulation results to validate the proposed reduced-order model for a representative wind farm composed of $8$ Type-3 wind turbines. We simulate and compare: i)~$N=8$ instances of the nonlinear state-space model in~\eqref{eq:nominal}, and ii)~one instance of the reduced-order model in~\eqref{eq:aggregate} using MATLAB's numerical solver ode45. Parameter values used in the simulations are listed in Appendix~B. Two disturbances are simulated simultaneously: i)~the input wind profile chosen is shown in Fig.~\ref{fig:plots}(a), and ii)~the grid voltage is given a step change at $10~[\mathrm{s}]$ as shown in Fig.~\ref{fig:plots}(b). Traces for a subset of electrical and mechanical state variables are shown in Figs.~\ref{fig:plots}(c)-(j). We express all quantities in $[\mathrm{p.u.}]$ on the individual turbine's base. As proved in Section~\ref{sec: ROM}, the voltages and turbine and generator speeds (Figs.~\ref{fig:plots}(c)-(f)) in the individual and reduced-order models are the same. However, the currents and powers in the aggregate model are scaled $8$ times (Fig.~\ref{fig:plots}(g)-(j)). The simulations are executed on a personal computer with an Intel Core $\mathrm i 5$ CPU and $8$ GB RAM. The computation time to run the model with $8$ turbines is $75.38 \mathrm{s}$, and that of the reduced-order model is $9.42 \mathrm{s}$. 

\section{Conclusions and Directions for Future Work} \label{sec:Conclusions}
An aggregate reduced-order structure-preserving model for wind farms composed of parallel connected Type-3 wind turbines was introduced. Parametric scalings were established and rigorously justified to validate the reduced-order model. Numerical simulations established the accuracy and computational benefits of the model. Ongoing work is focused on relaxing the assumption that turbine incident wind speeds are identical, and incorporating interconnecting impedances in the electrical network.

\section*{Appendix}
\subsection{Entries of state matrices $A$, $B$, and nonlinear function $g$}
Corresponding to the state-space model in~\eqref{eq:nominal}, we provide nonzero entries of matrices $A$ and $B$ in Tables~\ref{tbl:A_matrix}--\ref{tbl:B_matrix}, and then list the nonzero entries of function $g$. 
\begin{table}[h]\centering
\caption{Nonzero entries of matrix $A$}
\resizebox{1\linewidth}{!}{
  \begin{tabular}{| p{10mm} | p{63mm} | p{10mm} | p{60mm} |}
    \hline
    \emph{Entry} &     \emph{Value} &     \emph{Entry} &     \emph{Value} \\ \hline \hline
    (1,1) & $\omega_\mathrm{nom} (-R_1 + K_\mathrm{mrr}^2 k^{i\mathrm{d}}_\mathrm{RCC})/L_\mathrm{s}'$ & (1,2) & $-\omega_\mathrm{nom} \omega_\mathrm{s}$ \\ \hline
    (1,3) & $\omega_\mathrm{nom}K_\mathrm{mrr} k^{p\mathrm{d}}_{RCC} k^i_\mathrm{RPC}/L_\mathrm{s}'$  & (1,5) & $\omega_\mathrm{nom}K_\mathrm{mrr} k^{i\mathrm{d}}_\mathrm{RCC}/L_\mathrm{s}'$ \\ \hline
     (1,21) & $-\omega_\mathrm{nom}(K_\mathrm{mrr} k^{p\mathrm{d}}_\mathrm{RCC}/(L_\mathrm{s}' X_\mathrm{m}) + 1/(L_\mathrm{s}' T_\mathrm{r} \omega_\mathrm{s}))$ & (2,1) &  $\omega_\mathrm{nom} \omega_\mathrm{s}$ \\ \hline
     (2,2) & $\omega_\mathrm{nom} (-R_1 + K_\mathrm{mrr}^2 k^{i\mathrm{q}}_\mathrm{RCC})/L_\mathrm{s}'$ & (2,4) & $\omega_\mathrm{nom}K_\mathrm{mrr} k^{p\mathrm{q}}_\mathrm{RCC} k^i_\mathrm{RTC}/L_\mathrm{s}'$ \\ \hline
     (2,6) & $\omega_\mathrm{nom}K_\mathrm{mrr} k^{i\mathrm{q}}_\mathrm{RCC}/L_\mathrm{s}'$ & (2,20) & $\omega_\mathrm{nom}(-1/L_\mathrm{s}' T_\mathrm{r} \omega_\mathrm{s} + K_\mathrm{mrr} k^{p\mathrm{q}}_\mathrm{RCC}/(L_\mathrm{s}' X_\mathrm{m}))$ \\ \hline
     (5,1) & $K_\mathrm{mrr}$ & (5,3) & $k^i_\mathrm{RPC}$ \\ \hline
     (5,21) & $K_\mathrm{mrr}$ & (6,2) & $-1/X_\mathrm{m}$ \\ \hline
     (6,4) & $k^i_\mathrm{RTC}$ & (6,20) & $1/X_\mathrm{m}$ \\ \hline
     (7,7) & $-\omega_\mathrm{nom}k^p_\mathrm{GCC}/L_\mathrm{i}$ & (7,11) & $\omega_\mathrm{nom}k^i_\mathrm{GCC}/L_\mathrm{i}$ \\ \hline
     (7,14) & $-\omega_\mathrm{nom} k^p_\mathrm{GCC} k^p_\mathrm{GPC}/L_\mathrm{i}$ & (7,16) & $\omega_\mathrm{nom} k^p_\mathrm{GCC} k^i_\mathrm{GPC}/L_\mathrm{i}$ \\ \hline
     (7,22) & $-\omega_\mathrm{nom}/L_\mathrm{i}$ & (8,8) & $-\omega_\mathrm{nom} k^p_\mathrm{GCC}/L_\mathrm{i}$ \\ \hline
     (8,12) & $\omega_\mathrm{nom} k^i_\mathrm{GCC}/L_\mathrm{i}$ & (8,13) & $-\omega_\mathrm{nom}k^p_\mathrm{GCC} k^p_\mathrm{GPC}/L_\mathrm{i}$ \\ \hline
     (8,15) & $\omega_\mathrm{nom} k^p_\mathrm{GCC} k^i_\mathrm{GPC}/L_\mathrm{i}$ & (8,23) & $-\omega_\mathrm{nom}/L_\mathrm{i}$ \\ \hline
     (9,9) & $\omega_\mathrm{nom}$ & (9,22) & $\omega_\mathrm{nom}/L_\mathrm{g}$ \\ \hline
     (10,10) &  $\omega_\mathrm{nom}$ & (10,23) & $\omega_\mathrm{nom}/L_\mathrm{g}$ \\ \hline
     (11,1) & $-1$ & (11,14) & $-k^p_\mathrm{GPC}$ \\ \hline
     (11,16) & $k^i_\mathrm{GPC}$ &  (12,2) & $-1$ \\ \hline
     (12,13) & $-k^p_\mathrm{GPC}$ & (12,15) & $k^i_\mathrm{GPC}$ \\ \hline
     (13,13) & $-\omega_\mathrm{c,PC}$ & (14,14) & $-\omega_\mathrm{c,PC}$ \\ \hline
     (15,13) & $-1$ & (16,14) & $-1$ \\ \hline
     (17,17) & $-\omega_\mathrm{nom}c_\mathrm{sh}/(2H_\mathrm{g})$ & (17,18) & $\omega_\mathrm{nom}c_\mathrm{sh}/(2H_\mathrm{g})$ \\ \hline
     (17,19) & $k_\mathrm{sh}/(2H_\mathrm{g})$ & (18,17) & $\omega_\mathrm{nom}c_\mathrm{sh}/(2H_\mathrm{t})$ \\ \hline
     (18,18) & $-\omega_\mathrm{nom}c_\mathrm{sh}/(2H_\mathrm{t})$ & (18,19) & $-k_\mathrm{sh}/(2H_\mathrm{t})$ \\ \hline
	 (19,17) & $-\omega_\mathrm{nom}$ & (19,18)  & $\omega_\mathrm{nom}$ \\ \hline   
     (20,2) & $\omega_\mathrm{nom} \omega_\mathrm{s}(-R_2 + k^{p\mathrm{q}}_\mathrm{RCC} K_\mathrm{mrr}^2)$ & (20,4) & $\omega_\mathrm{nom} \omega_\mathrm{s} K_\mathrm{mrr} k^{p\mathrm{q}}_{RCC} k^i_\mathrm{RTC}$ \\ \hline
     (20,6) & $\omega_\mathrm{nom}\omega_\mathrm{s} K_\mathrm{mrr} k^{i\mathrm{q}}_\mathrm{RCC}$ & (20,20) & $\omega_\mathrm{nom}(-1/T_\mathrm{r} + \omega_\mathrm{s} K_\mathrm{mrr} k^{i\mathrm{q}}_\mathrm{RCC}/X_\mathrm{m})$ \\ \hline
     (20,21) & $-\omega_\mathrm{nom} \omega_\mathrm{s}$ & (21,4) &  $\omega_\mathrm{nom} \omega_\mathrm{s}(-R_2 + k^{p\mathrm{d}}_\mathrm{RCC} K_\mathrm{mrr}^2)$ \\ \hline
     (21,5) & $\omega_\mathrm{nom}\omega_\mathrm{s} K_\mathrm{mrr} k^{p\mathrm{d}}_\mathrm{RCC} k^i_\mathrm{RPC}$ & (21,7) & $\omega_\mathrm{nom}\omega_\mathrm{s} K_\mathrm{mrr} k^{i\mathrm{d}}_\mathrm{RCC}$ \\ \hline
     (21,20) & $\omega_\mathrm{nom}\omega_\mathrm{s}$ & (21,21) & $-\omega_\mathrm{nom}(1/T_\mathrm{r} + \omega_\mathrm{s} K_\mathrm{mrr} k^{p\mathrm{d}}_\mathrm{RCC}/X_\mathrm{m})$ \\ \hline
     (22,7) & $\omega_\mathrm{nom}/C_\mathrm{f}$ & (22,9) & $-\omega_\mathrm{nom}/C_\mathrm{f}$ \\ \hline
     (22,23) & $\omega_\mathrm{nom}$ & (23,8) & $\omega_\mathrm{nom}/C_\mathrm{f}$ \\ \hline
     (23,10) & $-\omega_\mathrm{nom}/C_\mathrm{f}$ & (23,22) &  $-\omega_\mathrm{nom}$ \\ \hline
     (24,24) &  $-\omega_\mathrm{c,PLL}$ & (25,24) & $-1$ \\ \hline
     (26,24) & $-k^p_\mathrm{PLL}$ & (26,25)  &  $k^i_\mathrm{PLL}$ \\ \hline
     (27,13) &  $-1/C$ &  & \\ \hline
  \end{tabular}
}
\label{tbl:A_matrix}
\end{table}

\begin{table}[h]\centering
\caption{Nonzero entries of matrix $B$}
\resizebox{0.85\linewidth}{!}{
  \begin{tabular}{| p{10mm} | p{40mm} | p{20mm} | p{40mm} |}
    \hline 
    \emph{Entry} & \emph{Value} & \emph{Entry} & \emph{Value} \\ \hline \hline
    (1,1) & $\omega_\mathrm{nom} K_\mathrm{mrr} k^{p\mathrm{d}}_\mathrm{RCC} k^p_\mathrm{RPC}/L_\mathrm{s}'$ & (3,1) & $1$ \\ \hline
    (5,1) & $k^p_\mathrm{RPC}$ & (21,1) & $-\omega_\mathrm{nom} \omega_\mathrm{s} K_\mathrm{mrr} k^{p\mathrm{d}}_\mathrm{RCC} k^\mathrm{p}_\mathrm{RPC}$ \\ \hline
    (7,1) & $\omega_\mathrm{nom} k_\mathrm{CC}^p k_\mathrm{PC}^p/L_\mathrm{i}$ & (11,1), (16,1) & $k^p_\mathrm{PC}$,\, $1$ \\ \hline
  \end{tabular}
}
\label{tbl:B_matrix}
\end{table}

The nonzero entries of $g(x,u_2,u_3)$ are given by
\begin{tiny}
\begin{align*}
&g_1 = \frac{\omega_\mathrm{nom} K_\mathrm{mrr} k^{p\mathrm{d}}_\mathrm{RCC} k^p_\mathrm{RPC}}{L_\mathrm{s}'} (v_{\mathrm{g}}^\mathrm{q} i_{\mathrm{s}}^\mathrm{d} - v_\mathrm{g}^\mathrm{d} i_{\mathrm{s}}^\mathrm{q} ) + \frac{\omega_\mathrm{nom}\omega_\mathrm{r}e_{\mathrm{s}}^\mathrm{d}}{L_\mathrm{s}'\omega_\mathrm{s}} - \frac{\omega_\mathrm{nom}}{L_\mathrm{s}'}v_\mathrm{g}^\mathrm{d}, \\
&g_2 =  -\frac{\omega_\mathrm{nom}K_\mathrm{mrr} k^{p\mathrm{q}}_\mathrm{RCC} k^p_\mathrm{RTC}}{L_\mathrm{s}'}\left(\frac{v_{\mathrm{g}}^\mathrm{q}}{\omega_\mathrm{s}} i_{\mathrm{s}}^\mathrm{q} + \frac{v_{\mathrm{g}}^\mathrm{d}}{\omega_\mathrm{s}} i_{\mathrm{s}}^\mathrm{d} - K_\mathrm{opt}\omega_\mathrm{r}^2\right) + \frac{\omega_\mathrm{nom} \omega_\mathrm{r}e_{\mathrm{s}}^\mathrm{q}}{L_\mathrm{s}'\omega_\mathrm{s}} \\
&\quad   - \frac{\omega_\mathrm{nom}}{L_\mathrm{s}'}v_\mathrm{g}^\mathrm{d},\,
g_3 = v_{\mathrm{g}}^\mathrm{q} i_{\mathrm{s}}^\mathrm{d} - v_{\mathrm{g}}^\mathrm{d} i_{\mathrm{s}}^\mathrm{q},\, g_4 = K_\mathrm{opt} \omega_\mathrm{r}^2-\left(\frac{v_{\mathrm{g}}^\mathrm{q}}{\omega_\mathrm{s}} i_{\mathrm{s}}^\mathrm{q} + \frac{v_{\mathrm{g}}^\mathrm{d}}{\omega_\mathrm{s}} i_{\mathrm{s}}^\mathrm{d}\right),\\
&g_5 = k^p_\mathrm{RPC} (v_{\mathrm{g}}^\mathrm{q} i_{\mathrm{s}}^\mathrm{d} - v_{\mathrm{g}}^\mathrm{d} i_{\mathrm{s}}^\mathrm{q}),\, g_6 = k^p_\mathrm{RTC}\left(K_\mathrm{opt} \omega_\mathrm{r}^2-\frac{v_{\mathrm{g}}^\mathrm{q}}{\omega_\mathrm{s}} i_{\mathrm{s}}^\mathrm{q} - \frac{v_{\mathrm{g}}^\mathrm{d}}{\omega_\mathrm{s}} i_{\mathrm{s}}^\mathrm{d}\right), \\
&g_8 = \omega_\mathrm{nom} \frac{k^p_\mathrm{CC} k^p_\mathrm{PC}}{L_\mathrm{i}} g_{15},\, g_9 = \omega_\mathrm{nom} ((-k^p_{\text{PLL}} v_{\mathrm{PLL}} + k^i_{\text{PLL}}  \phi_{\text{PLL}}) i_{\mathrm{g}}^\mathrm{q} - \frac{1}{L_\mathrm{g}} v_\mathrm{g}^\mathrm{d}), 
\\
&g_{10} = \omega_\mathrm{nom} ((k^p_{\text{PLL}} v_{\mathrm{PLL}} - k^i_{\text{PLL}}  \phi_{\text{PLL}}) i_{\mathrm{g}}^\mathrm{d} - \frac{1}{L_\mathrm{g}} v_\mathrm{g}^\mathrm{q}), 
\\
&g_{13} = \omega_{\mathrm{c,PC}} \left( v_\mathrm{g}^\mathrm{d} i_\mathrm{g}^\mathrm{d} + v_\mathrm{g}^\mathrm{q} i_\mathrm{g}^\mathrm{q}\right),\, 
g_{14} = \omega_{\mathrm{c,PC}} \left(-v_\mathrm{g}^\mathrm{q} i_\mathrm{g}^\mathrm{d} + v_\mathrm{g}^\mathrm{d} i_\mathrm{g}^\mathrm{q} \right), \\
&g_{15} = \left( k^{p\mathrm{d}}_\mathrm{RCC}(k^p_\mathrm{RPC}(q^* - (-v_\mathrm{g}^\mathrm{q} i_\mathrm{s}^\mathrm{d} + v_\mathrm{g}^\mathrm{d} i_\mathrm{s}^\mathrm{q})) + k^i_\mathrm{RPC} \phi_\mathrm{r}^\mathrm{q} - i_\mathrm{r}^\mathrm{d} + k^{i\mathrm{d}}_\mathrm{RCC} \phi_\mathrm{r}^\mathrm{id} \right) \left( \frac{e_\mathrm{s}^\mathrm{q}}{X_\mathrm{m}} \right.\\
&\left. - K_\mathrm{mrr}i_\mathrm{s}^\mathrm{d} \right) + \left( k^{p\mathrm{q}}_\mathrm{RCC}(k^p_\mathrm{RPC}(K_\mathrm{opt}\omega_\mathrm{r}^2 - (\frac{e_\mathrm{s}^\mathrm{d}}{\omega_\mathrm{s}} i_\mathrm{s}^\mathrm{d} + \frac{e_\mathrm{s}^\mathrm{q}}{\omega_\mathrm{s}} i_\mathrm{s}^\mathrm{q})) + k^i_\mathrm{RPC} \phi_\mathrm{r}^\mathrm{t} - i_\mathrm{r}^\mathrm{q}) \right. \\
&\left. + k^{i\mathrm{q}}_\mathrm{RCC} \phi_\mathrm{r}^\mathrm{iq} \right) \left( \frac{e_\mathrm{s}^\mathrm{d}}{X_\mathrm{m}} - K_\mathrm{mrr}i_\mathrm{s}^\mathrm{q} \right), \\
&g_{17} = -\frac{1}{2H_\mathrm{g}}\left( \frac{\mathrm{e}_\mathrm{s} ^ \mathrm{q}}{\omega_\mathrm{s}}i_\mathrm{s}^\mathrm{q} + \frac{\mathrm{e}_\mathrm{s}^\mathrm{d}}{\omega_\mathrm{s}}i_\mathrm{s}^\mathrm{d} \right),\, g_{18} = \frac{1}{2T_{\mathrm{m,base}}}\left(\rho \pi R^2 C_p(\lambda,\beta)\frac{v_{\mathrm{w}}^3}{\omega_\mathrm{t}}\right), \\
&g_{20} = -\omega_\mathrm{nom} \omega_\mathrm{s} K_\mathrm{mrr} k^{p\mathrm{q}}_\mathrm{RCC} k^p_\mathrm{RTC}\left(\frac{v_{\mathrm{g}}^\mathrm{q}}{\omega_\mathrm{s}} i_{\mathrm{s}}^\mathrm{q} + \frac{v_{\mathrm{g}}^\mathrm{d}}{\omega_\mathrm{s}} i_{\mathrm{s}}^\mathrm{d} - K_\mathrm{opt}\omega_\mathrm{2}^2 \right) + \omega_{\mathrm{nom}} \omega_\mathrm{r}  e_{\mathrm{s}}^\mathrm{q},\\
&g_{21} = -\omega_\mathrm{nom} \left( \omega_\mathrm{s} K_\mathrm{mrr} k^{p\mathrm{d}}_\mathrm{RCC} k^p_\mathrm{RPC} (v_{\mathrm{g}}^\mathrm{q} i_{\mathrm{s}}^\mathrm{d} - v_{\mathrm{g}}^\mathrm{d} i_{\mathrm{s}}^\mathrm{q})  -\omega_\mathrm{r}  e_{\mathrm{s}}^\mathrm{d} \right), \\
&g_{22} = \omega_\mathrm{nom} (-k^p_{\text{PLL}} v_{\mathrm{PLL}} + k^i_{\text{PLL}}  \phi_{\text{PLL}})  v_\mathrm{f}^\mathrm{q}, \\
&g_{23} = \omega_\mathrm{nom} (k^p_{\text{PLL}} v_{\mathrm{PLL}} - k^i_{\text{PLL}}  \phi_{\text{PLL}})  v_\mathrm{f}^\mathrm{d},\,
g_{24} = \omega_{\mathrm{c,PLL}} v_\mathrm{g}^\mathrm{d},\,
g_{26} = 1, g_{27} = \frac{1}{C} g_{15},
\end{align*}
\end{tiny}
where  $v_\mathrm{g}^\mathrm{q}$ and $v_\mathrm{g}^\mathrm{d}$ are given by
\begin{tiny}
\begin{align*}
v_\mathrm{g}^\mathrm{q} &= \frac{2}{3}\left(\cos(\delta)v_\mathrm{g}^\mathrm{a}+\cos(\delta-\frac{2\pi}{3})v_\mathrm{g}^\mathrm{b} + \cos(\delta+\frac{2\pi}{3})v_\mathrm{g}^\mathrm{c}\right), \\
v_\mathrm{g}^\mathrm{d} &= -\frac{2}{3} \left(\sin(\delta)v_\mathrm{g}^\mathrm{a}+\sin(\delta-\frac{2\pi}{3})v_\mathrm{g}^\mathrm{b}  + \sin(\delta+\frac{2\pi}{3})v_\mathrm{g}^\mathrm{c}\right). 
\end{align*}
\end{tiny}

\subsection{List of Parameters and Numerical Values}
In Table~\ref{tbl:PDescription}, we list pertinent parameters that characterize the wind turbine models, and provide numerical values that are used in the simulations. For the aggregate model, pertinent parameters are scaled with $N=8$ (see~\eqref{eq:scalings}). 
\begin{table}[h]\centering
\caption{Parameter Description}
\resizebox{0.85\linewidth}{!}{
  \begin{tabular}{| p{20mm} | p{40mm} | p{25mm} | }
  \hline
  \emph{Parameter} & \emph{Description} & \emph{Value} \\ \hline \hline
  $\omega_{\mathrm{nom}}$ &  Synchronous Base Speed & $2\pi60$\\ \hline
  $\omega_\mathrm{s}$  & Synchronous Speed & $1$ $[\mathrm{p.u.}]$\\ \hline
  $L_\mathrm{m}$ &  Mutual Inductance & $4$ $[\mathrm{p.u.}]$\\ \hline
   $L_\mathrm{s}$ &  Stator Inductance & $1.101L_\mathrm{m}$ \\ \hline
  $L_\mathrm{r}$ & Rotor Inductance & $1.005L_{\mathrm{s}}$\\  \hline
  $K_{\mathrm{mrr}}$ & Mutual Coupling Factor & $L_\mathrm{m}/L_\mathrm{r}$ \\ \hline
  $L_\mathrm{s}'$ & Inductance referred to the stator & $L_\mathrm{{s}} - L_{\mathrm{m}}K_{\mathrm{mrr}}$  \\ \hline
  $R_\mathrm{s}$ & Stator Resistance & $0.005$ $[\mathrm{p.u.}]$ \\  \hline
   $R_\mathrm{r}$ & Rotor Resistance & $1.1R_\mathrm{s}$  \\ \hline 
  $R_1$ & Resistance referred to the stator & $R_\mathrm{s} + R_2$  \\ \hline
  $R_2$ & Resistance referred to the rotor &$K_\mathrm{mrr} ^2 R_\mathrm{r}$ \\ \hline
$T_\mathrm{r}$ &Rotor Time Constant& $L_\mathrm{r}/R_\mathrm{r}$  \\ \hline
$H_\mathrm{t}$ &  Turbine Inertia & $4$ $[\mathrm{s}]$\\ \hline
 $H_\mathrm{g}$ & Generator Inertia & $0.1H_\mathrm{t}$\\ \hline
$k_{\mathrm{sh}}$ & Drive Train Shaft Stiffness & $0.3$ $[\mathrm{p.u./el.rad}]$\\ \hline
 $c_{\mathrm{sh}}$ & Drive Train Damping Coefficient & $0.01$ $[\mathrm{p.u.s/el.rad}]$\\ \hline
$\beta$ &  Blade Pitch Angle & $0 ^{\circ}$\\ \hline
$C_{p,\mathrm{max}}$ & Maximum value of $C_p$  & $0.4382$ \\ \hline
$P_\mathrm{rated}$ & Rated power of the turbine & $5$ $[\mathrm{MW}]$ \\ \hline
$R$ & Blade Length & $58.6$ $[\mathrm{m}]$ \\ \hline
$\rho$ & Air Density & $1.225$ $[\mathrm{kg/m^3}]$ \\ \hline

   \end{tabular}
}
\label{tbl:PDescription}
\end{table}

\bibliographystyle{ieeetr}
\bibliography{references}
\end{document}